\documentclass[aps,showpacs,tightenlines,epsfig]{revtex4}
\usepackage{graphicx}
\usepackage{ams}
\begin{document}
\title{Quantum random walk of the field in an externally driven cavity}
\author{G. S. Agarwal$^*$ and P. K. Pathak\footnote{On leave:
Physical Research Laboratory, Navrangpura, Ahmedabad-380 009,
India}}
\address{Department of Physics, Oklahoma State University, Stillwater,
OK-74078}
\date{\today}
\begin{abstract}
Using resonant interaction between atoms and the field in a high
quality cavity, we show how to realize quantum random walks as
proposed by Aharonov et al [Phys. Rev. A {\bf48}, 1687 (1993)]. The
atoms are driven strongly by a classical field. Under conditions of
strong driving we could realize an effective interaction of the form
$ iS^{x}(a-a^{\dag})$ in terms of the spin operator associated with
the two level atom and the field operators. This effective
interaction generates displacement in the field's wavefunction
depending on the state of the two level atom. Measurements of the
state of the two level atom would then generate effective state of
the field. Using a homodyne technique, the state of the quantum
random walker can be monitored.
\end{abstract}
\pacs{42.50.Pq, 03.67.Lx}
\maketitle
\section{Introduction}
In a very interesting paper Aharonov et al \cite{Aharonov} proposed
the idea of a quantum random walk. Here a random walker is
constrained to move left or right depending on the state of an
auxiliary quantum mechanical system. One then examines the state of
the random walker subject to the measurement of the state of the
auxiliary system. As an interesting consequence of this quantum
random walk, Aharonov et al \cite{Aharonov} found that the walker's
distribution could shift by an amount which could be larger than the
width of the initial distribution. Further the displacement could be
much larger than the classical displacement. Several proposals
\cite{Cavity,Aharonov,Sanders,milburn,lattice,exp,knight,bouwmeester}
exist for realizations of the quantum random walk. For example
Aharonov et al gave a cavity QED model where the photon number
distribution can get displaced. Sanders et al \cite{Sanders}
considered a dispersive interaction in the cavity of the form $
S^{z}(a+a^{\dag})$ and considered the random walk of the field on
states on a circle. Other interesting theoretical schemes for
implementing quantum walks have been suggested in ion-traps
\cite{milburn} and in optical lattices \cite{lattice}. Knight et al
\cite{knight} further showed that an earlier experiment
\cite{bouwmeester} was a realization of quantum random walks. A
scheme using linear optical elements has been recently implemented
\cite{exp}.

Here we propose a method which yields precisely quantum random walk
as proposed by Aharonov et al. We use cavity QED however we drive
the atoms by an external field. Currently there is considerable
progress in realizing a variety of high quality cavities and a
variety of interactions  and thus one is in a situation where
proposals like the one presented here are likely to be implemented.

The organization of the paper is as follows. In Sec.II we present
the details of our model and show the conditions under which such a
model gives rise to an effective Hamiltonian which we use in Sec.III
to realize quantum random walk. In this section we also present the
results for the Wigner function for the state of the quantum walker.
In Sec.IV we show how the homodyne measurements of the field can be
used to check the characteristics of the quantum random walk. In
Sec.V we incorporate the effects of decoherence due to the decay of
the field in the cavity. In the appendix we discuss the state of the
walker if no conditional measurements are made and establish
relation to classical random walks.
\section{Effective hamiltonian for Quantum Random Walk using driven atoms}
We consider a two level Rydberg atom having its higher energy state
$|e\rangle$ and lower energy state $|g\rangle$, interacting with a
single mode of the electromagnetic field in a cavity. The atom
passes through the cavity and interacts resonantly with the field.
Further the atom is driven by a strong classical field. For
simplicity we choose atomic transition frequency, the cavity
frequency and the frequency of the driving field to be same. The
Hamiltonian for the system in the interaction picture is written as
\begin{equation}
H=-i\hbar g\left(S^{+} a-a^{\dag}S^{-}\right)+ \hbar\left(S^{+}
{\cal{E}} +S^{-}{\cal{E}}^{*}\right),\label{ham}
\end{equation}
where $g$ and ${\cal{E}}$ are the coupling constants of the
interaction of the atom with the cavity field and with the deriving
field. We have chosen $g$ as real and ${\cal{E}}$ as complex. The
annihilation (creation) operator for the field in the cavity is $a
(a^{\dag})$ and $S^{+},~S^{-}$ are atomic spin operators. The last
term in Eq.(\ref{ham}) is the interaction with the external field.
We further rewrite the above Hamiltonian in a picture in which the
interaction with the external field has already been diagonalized.
\begin{equation}
|\bar{\psi}\rangle=e^{iht}|\psi\rangle;~h=S^+{\cal{E}}+S^-{\cal{E}}^*,
\label{h}
\end{equation}
where $|\bar{\psi}\rangle$ is transformed atomic state in new
picture from old atomic state $|\psi\rangle$. The Hamiltonian in
this picture is
\begin{eqnarray}
\label{newH}
&&\bar{H}=-ige^{iht}(S^+a-S^-a^{\dag})e^{-iht},\\
\label{trnsf} &&e^{iht}\equiv\cos(|{\cal{E}}|t)+\frac{i
h}{|{\cal{E}}|}\sin(|{\cal{E}}|t).
\end{eqnarray}
The atomic spin operators $S^{\pm}$ transform as
\begin{eqnarray}
\label{s+} e^{iht}S^{+}e^{-iht}\equiv
S^+\cos^2(|{\cal{E}}|t)+\frac{{\cal{E}}^{*2}}
{|{\cal{E}}|^2}\sin^2(|{\cal{E}}|t)S^{-}\nonumber\\-\frac{2i{\cal{E}^{*}}}{|{\cal{E}}|}
S^{z}\sin(|{\cal{E}}|t)\cos(|{\cal{E}}|t),\\
\label{s-} e^{iht}S^{-}e^{-iht}\equiv
S^{-}\cos^2(|{\cal{E}}|t)+\frac{{\cal{E}}^{2}}
{|{\cal{E}}|^2}\sin^2(|{\cal{E}}|t)S^{+}\nonumber\\+\frac{2i{\cal{E}}}{|{\cal{E}}|}
S^{z}\sin(|{\cal{E}}|t)\cos(|{\cal{E}}|t).
\end{eqnarray}
Using Eqs.(\ref{s+}) and (\ref{s-}), Eq.(\ref{newH}) becomes
\begin{eqnarray}
\bar{H}=-ig\left(S^+\cos^2(|{\cal{E}}|t)+\frac{{\cal{E}}^{*2}}
{|{\cal{E}}|^2}\sin^2(|{\cal{E}}|t)S^{-}\right.\nonumber\\
\left.-\frac{2i{\cal{E}^{*}}}{|{\cal{E}}|}
S^{z}\sin(|{\cal{E}}|t)\cos(|{\cal{E}}|t)\right)a-H.c. \label{Hbar}
\end{eqnarray}
We note that the Hamiltonians of the above form have been previously
used to treat the inhibition of the spontaneous emission \cite{spe}
and for the production of Schrodinger cat states \cite{Solano}. We
assume that the atom is driven strongly so that $|{\cal{E}}|$ is
large and hence we drop rapidly oscillating terms from
Eq.(\ref{Hbar}) {\it i.e.} $e^{\pm2i|{\cal{E}}|t}\Rightarrow0$. Then
Eq.(\ref{Hbar}) reduces to
\begin{equation}
\bar{H}=-\frac{ig}{2}\left(S^++\frac{{\cal{E}}^{*2}}{|{\cal{E}}|^2}S^{-}\right)a-H.c.
\label{newHbar}
\end{equation}
We choose ${\cal{E}}^{*2}/{|\cal{E}}|^2=1$, in general, this can
also be done by adjusting phases with atomic operators. Then the
Eq.(\ref{newHbar}) takes the form
\begin{equation}
\bar{H}_{eff}=gS^{x}\left(\frac{a-a^{\dag}}{i}\right). \label{netH}
\end{equation}
Note the appearance of the well known displacement
$D(\alpha)=(a^{\dag}\alpha-a\alpha^*)$ in the Eq.(\ref{netH}). In
particular we have the momentum operator (out of phase quadrature
for the field). Further it should also be noted that $h$ as defined
by Eq.(\ref{h}) commutes with $\bar{H}_{eff}$. In the original
interaction picture the Hamiltonian for our model will be
\begin{equation}
H_{eff}=gS^{x}\left(\frac{a-a^{\dag}}{i}\right)+2|{\cal{E}}|S^{x}.
\label{heff}
\end{equation}
In the effective Hamiltonian (\ref{heff}) field displacement
operator appears with atomic operator, which can produce
displacement in field state depending on the atomic state.
\section{realization of random walk}
\begin{figure}
\includegraphics[width=3in, height=3in]{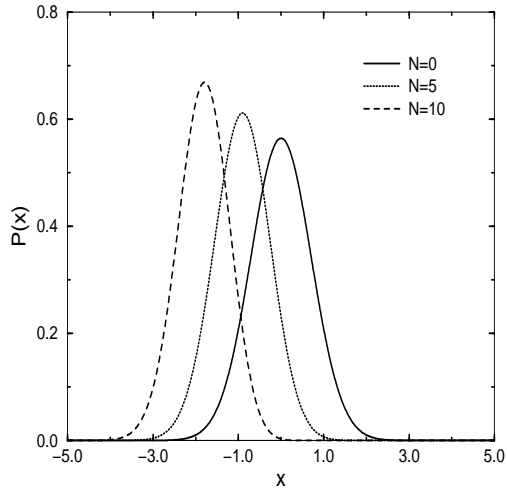}
\caption{The probability distribution $P(x)$ for the position of the
quantum random walker, assuming initial wave packet as Gaussian
$\exp[-(x-\alpha)^2/2]$ for $\alpha=0$, step size $l=0.05$,
$\phi=2\pi$ and $\theta=2\pi/3$.} \label{fig1}
\end{figure}

We next examine the evolution of the system of the two level atom
and the field inside the cavity. Let us consider that, initially the
atom is in the superposition state
$|\Phi\rangle=(c_1|e\rangle+c_2|g\rangle)$ and the field is in a
coherent state $|\alpha\rangle$. Using Eq.(\ref{heff}) the combined
state of the atom-cavity system after time $t$ is given by
\begin{eqnarray}
|\psi(t)\rangle&=&\exp\left[gtS^{x}(a^{\dag}-a)-2i|{\cal{E}}|tS^{x}\right]|\Phi\rangle|\alpha\rangle,\\
&=&\frac{c_+e^{-i\phi}}{2}\left(|g\rangle+|e\rangle\right)|\alpha+gt/2\rangle\nonumber\\
&+&\frac{c_-e^{i\phi}}{2}\left(|g\rangle-|e\rangle\right)|\alpha-gt/2\rangle,\\
\label{app}
&=&|g\rangle \left[\frac{c_+e^{-i\phi}}{2}|\alpha+gt/2\rangle+\frac{c_-e^{i\phi}}{2}|\alpha-gt/2\rangle\right]\nonumber\\
&+&|e\rangle\left[\frac{c_+e^{-i\phi}}{2}|\alpha+gt/2\rangle-\frac{c_-e^{i\phi}}{2}|\alpha-gt/2\rangle\right];\\
\phi&=&\left(|{\cal{E}}|+\frac{g}{2}Im(\alpha)\right)t;
\end{eqnarray}
where $c_+=c_1+c_2$ and $c_-=c_1-c_2$. Using normalization of atomic
states we can select $c_-/c_+=\tan\theta$. Thus the detection of the
atom in state $|e\rangle$ or $|g\rangle$ leaves the cavity field in
a superposition of states $|\alpha+gt/2\rangle$ and
$|\alpha-gt/2\rangle$. For small values of $gt$ the states
$|\alpha+gt/2\rangle$ and $|\alpha-gt/2\rangle$ overlap completely
and thus quantum interference effects between $|\alpha+gt/2\rangle$
and $|\alpha-gt/2\rangle$ becomes significant. If we assume that the
atom is detected in its ground state $|g\rangle$. Then the state of
the field inside the cavity can be written as
\begin{eqnarray}
|\psi_f\rangle\propto\left[e^{-i|{\cal{E}}|t}D(gt/2)+e^{i|{\cal{E}}|t}\tan(\theta)D(-gt/2)\right]|\alpha\rangle,
\end{eqnarray}
Clearly after passing one atom through the cavity the field inside
the cavity is displaced backward or forward along the line in a
random way by the step of $gt/2$. We can now iterate the above step
to obtain the state of the field after the passage of $N$ atoms. We
assume that atoms enter in the cavity in the state $|\Phi\rangle$
and after interaction with the field inside the cavity detected in
their ground state $|g\rangle$. Note that the displacement operators
appearing in the above state commute each other
$[D(gt/2),D(-gt/2)]=0$ for real $gt$. Thus the field state after the
passage of $N$ atoms is given by
\begin{eqnarray}
|\psi_f(N)\rangle&=&
C\left[e^{-i|{\cal{E}}|t}D(gt/2)+e^{i|{\cal{E}}|t}\tan(\theta)D(-gt/2)\right]^N|\alpha\rangle,\nonumber\\
&=&C\sum_{m=0}^{N} \left(
                    \begin{array}{c}
                      N \\
                      m \\
                    \end{array}
                  \right)\left[e^{-im|{\cal{E}}|t}D^m\left(\frac{gt}{2}\right)\times\right.~~~~~~~~\nonumber\\
&&\left.e^{i(N-m)|{\cal{E}}|t}
(\tan\theta)^{N-m}D^{N-m}\left(-\frac{gt}{2}\right)\right]|\alpha\rangle,\nonumber\\
&=&C\sum_{m=0}^{N} \left(
                    \begin{array}{c}
                      N \\
                      m \\
                    \end{array}
                  \right)
e^{i(N-2m)|{\cal{E}}|t}(\tan\theta)^{N-m}\nonumber\\
&&D^{N-2m}(-gt/2)|\alpha\rangle,\\
&=&C\sum_{m=0}^{N} \left(
                    \begin{array}{c}
                      N \\
                      m \\
                    \end{array}
                  \right)
e^{i(N-2m)\phi}(\tan\theta)^{N-m}\nonumber\\
&&|\alpha-(N-2m)gt/2)\rangle, \label{final}
\end{eqnarray}
where $C$ is normalization constant and we have used the property of
the displacement operator $D^{-1}(\alpha)=D(-\alpha)$. On writing
the above result in coordinate space representation, we get the
wavefunction $\psi_N(x,\alpha)=\langle x|\psi_f(N)\rangle$
\begin{eqnarray}
\psi_N(x,\alpha)=C\sum_{m=0}^{N} \left(
                    \begin{array}{c}
                      N \\
                      m \\
                    \end{array}
                  \right)
e^{i(N-2m)\phi}(\tan\theta)^{N-m}\nonumber\\\psi_{\alpha}\left(x+[N-2m]l\right),\label{wave}
\end{eqnarray}
where $\psi_{\alpha}(x)\equiv\langle x|\alpha\rangle$ is the
wavefunction corresponding to the initial cavity field state
$|\alpha\rangle$ which is centered at $x=\alpha$  and the step size
of the random walker is $l=gt/2$. We note that we have recovered the
result of Aharonov et al \cite{Aharonov}. In Fig.\ref{fig1} we have
plotted the probability amplitude distribution for initial wave
function $\psi_{\alpha}(x)\sim \exp[-(x-\alpha)^2/2]$ for real
values of $x$ and $\alpha=0$. The displacement depends on $\theta$,
$\phi$ and the number of steps $N$. The unexpected displacement in
the state of the random walker is the result of constructive quantum
interference between the states generated in various steps which
comes from the off diagonal terms in $P(x)=|\psi_N(\alpha,x)|^2$. We
have checked this by dropping the off diagonal terms in $P(x)$, in
that case $P(x)$ remains same in shape as the initial wave packet
but shifts by an amount $Nl$. The displacement of the random walker
is not bounded by the classically possible maximum and minimum
displacements $\pm Nl$. The quantum interference leads to an
arbitrary displacement in the random walker's position and can be
much larger than $\pm Nl$. A small squeezing in wavepacket is also
generated from these interference effects. The selection of phase
$\phi$ is also critical for displacement in the position of quantum
walker, for example for the parameters used in Fig.\ref{fig1} the
maximum displacement in the position of quantum walker occurs when
$\phi$ is integer multiple of $\pi$ and there will be minimum
displacement when $\phi$ is half integer multiple of $\pi$.

\begin{figure}
\includegraphics[width=5in, height=6in]{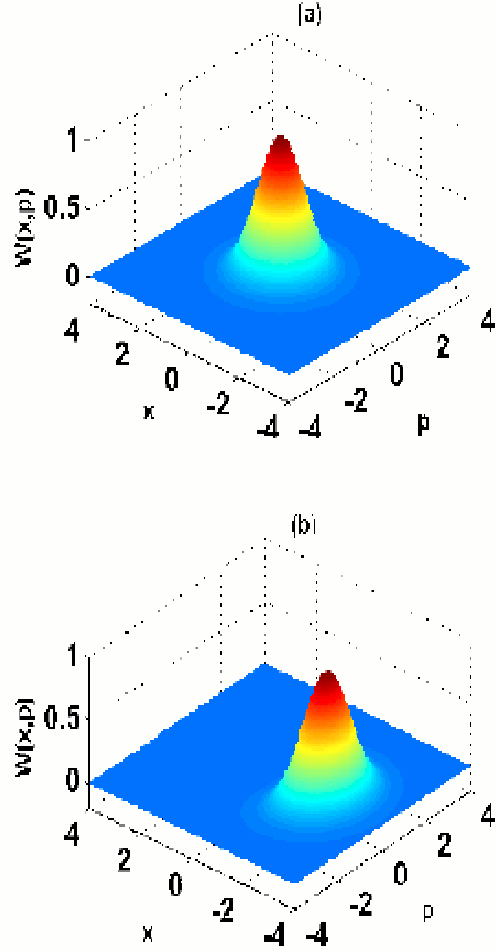}
\caption{The Wigner function $W(x,p)$ of the state of the random
walker, after number of steps (a) $N=0$ (b) $N=10$, using same
parameters as in FIG.\ref{fig1}.} \label{fig2}
\end{figure}

For visualizing quantum interferences we plot the Wigner function of
the random walker in Fig.\ref{fig2}. The Wigner distribution for any
state $\psi(x)$ can be obtained by using the definition
\cite{wigner},
\begin{equation}
W(x,p)=\frac{1}{\pi\hbar}\int e^{2ipy/\hbar}
\psi(x-y)\psi^{*}(x+y)dy.
\end{equation}
In the Fig.\ref{fig2}(a) the field is in its initial coherent state
and the wigner function is perfect Gaussian. As the field is
displaced by random steps, by passing atoms through the cavity,
quantum interference effects start deforming the shape of the Wigner
function from the Gaussian. After few steps the Wigner function is
squeezed in $x$ quadrature and gets displaced by an arbitrary
distance in $x$. In Fig.\ref{fig2}(b), (see also Fig.\ref{fig4}(a)),
we have shown the Wigner function after $10$ random steps for
initial Gaussian wave packet. The squeezing is also clear from the
Fig.\ref{fig1} which shows the narrowing of the distribution $P(x)$.
It is clear that the displacement in the position of random walker
comes as a result of quantum interference which is consequence of
quantum coherence between the states generated in random steps.
\section{measurement of the state of the random walker}
\begin{figure}
\centering
\includegraphics[width=3in]{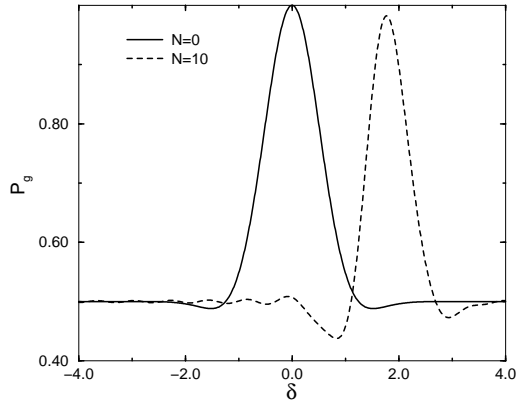}
\caption{The probability of detecting probe atom in its ground state
as a function of $\delta$ for the state of the quantum random walker
after number of steps $N=0$ ( solid line) and $N=10$ (dashed line).
The parameters used are same as in Fig.\ref{fig1} and the
interaction time for the probe atom is selected such that $gt_p=1.5
\pi$.} \label{fig3}
\end{figure}

\begin{figure}[h]
\includegraphics[width=5in, height=5in]{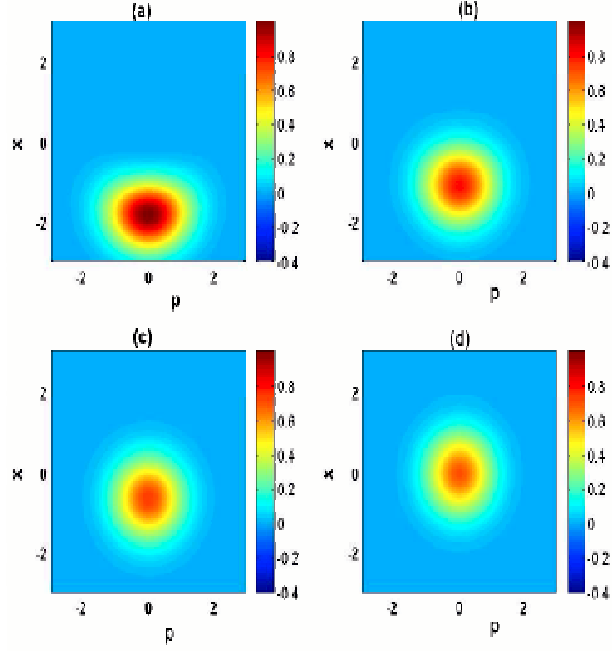}
\caption{The decoherence of the state of the random walker in terms
of Wigner function at different times, (a) for $\kappa t=0$, (b) for
$\kappa t=1/4N^2l^2$, (c) for $\kappa t=1/2N^2l^2$, (d) for $\kappa
t =2/N^2l^2$, other parameters are same as FIG.\ref{fig2} (b).}
\label{fig4}
\end{figure}

We next discuss how we can probe the quantum state of the random
walker. We propose homodyne techniques \cite{homodyne} for measuring
the state of the random walker. Such homodyne measurement can be
performed by mixing an external resonant coherent field to the
cavity and then probing the resultant cavity field by passing a test
atom through the cavity. In the previous section, we have shown how
the cavity field is displaced backward or forward in a random step
by passing single atom through the cavity. The state of the field in
the cavity after such $N$ steps can be monitored by homodyne
measurements which can be implemented in the same experimental set
up. After displacing the field inside the cavity by $N$ random
steps, by passing $N$ atoms, a resonant external coherent field
$|\beta\rangle$ is injected into the cavity. After adding the
external field, the state of the resultant field in the cavity is
\begin{eqnarray}
|\psi_H\rangle&=&C\sum_{m=0}^{N} \left(
                    \begin{array}{c}
                      N \\
                      m \\
                    \end{array}
                  \right)
e^{i(N-2m)\phi}(\tan\theta)^{N-m}\nonumber\\
&&D(\beta)|\alpha-(N-2m)gt/2)\rangle,\nonumber\\
&=&C\sum_n\sum_{m=0}^{N} \left(
                    \begin{array}{c}
                      N \\
                      m \\
                    \end{array}
                  \right)
e^{i(N-2m)\phi}(\tan\theta)^{N-m}\nonumber\\
&&\langle n|D(\beta)|\alpha-(N-2m)gt/2)\rangle|n\rangle,\nonumber\\
\label{disp}
&=&\sum_n F_n|n\rangle\\
\label{fm} F_n&=&C\sum_{m=0}^{N} \left(
                    \begin{array}{c}
                      N \\
                      m \\
                    \end{array}
                  \right)
e^{i(N-2m)\phi}(\tan\theta)^{N-m}\nonumber\\
&&\langle n|D(\beta)|\alpha-(N-2m)gt/2)\rangle.
\end{eqnarray}
Now we bring a similar atom in its lower energy state $|g\rangle$ to
probe the cavity field. The probability of detecting the probe atom
in its lower state $|g\rangle$ after crossing the cavity in time
$t_p$ is
\begin{equation}
P_g=\sum_n|F_n|^2\cos^{2}(gt_p\sqrt{n}).
\end{equation}
The interaction time $t_p$ for the probe atom is selected such that
if there are photons in the cavity it leaves the cavity in its
higher energy state $|e\rangle$ with larger probability. If we
choose the external field $|\beta\rangle$ such that
$\beta=-\alpha+\delta$, the probe atom will leave the cavity in its
ground state with larger probability when the value of $\delta$ will
be opposite and equal to the displacement of the random walker from
the initial position $\alpha$. Thus the probability of the probe
atom leaving the cavity in its lower state $|g\rangle$ would, as a
function of $\delta$, have peak corresponding to the positions of
the random walker after $N$ steps. In Fig.\ref{fig3}, we plot the
probability of detecting the probe atom in its lower state with
$\delta$. The solid line curve is result of homodyne measurement of
the position of the random walker corresponding to its initial
state. The dashed line curve is corresponding to the homodyne
measurement after $10$ steps using the same parameter as in
Fig.\ref{fig1}. Clearly the homodyne measurement yields the state of
the quantum walker (Fig.\ref{fig1}). Thus the homodyne measurement
can be an elegant way for monitoring the position of the random
walker in our model of realizing quantum random walks.

\section{decoherence of the generated state of the random walker}
Quantum random walks are different from the classical random walks
in the sense of quantum interferences which may lead much larger
displacements in the position of quantum random walker than the
classically possible maximum displacements. These quantum
interferences are the consequences of coherence in the system.
Clearly we need the coherence to live for a long time and thus it is
important to study the effects of the decoherence of the system. In
this section we study the decoherence of the state of the random
walker due to damping in the cavity. This can be done using the
master equation
\begin{equation}
\dot{\rho}=-\frac{\kappa}{2}(a^{\dag}a\rho-2a\rho a^{\dag}+\rho
a^{\dag}a),
\end{equation}
where $\kappa$ is cavity field decay parameter and we carry analysis
in the absence of thermal photons. For initial state (\ref{final})
we find the density matrix after time $t$

\begin{eqnarray}
\rho(t)&=&|C|^2\sum_{m=0}^{N}\sum_{n=0}^{N}\left(
                    \begin{array}{c}
                      N \\
                      m \\
                    \end{array}
                  \right)\left(
                    \begin{array}{c}
                      N \\
                      n \\
                    \end{array}
                  \right)
e^{2i(n-m)\phi}(\tan\theta)^{2N-m-n}\nonumber\\
&&\langle \alpha-(N-2m)l|\alpha-(N-2n)l)\rangle^{(1-e^{-\kappa t})}\nonumber\\
&&|\alpha-(N-2m)l\rangle_t\langle\alpha-(N-2n)l|_t~, \label{deco}
\end{eqnarray}
where $|\zeta\rangle_t\equiv|\zeta e^{-\kappa t/2}\rangle$. In the
limit $\kappa t<<1$ the Eq.(\ref{deco}) simplifies to
\begin{eqnarray}
\rho(t)&=&|C|^2\sum_{m=0}^{N}\sum_{n=0}^{N}\left(
                    \begin{array}{c}
                      N \\
                      m \\
                    \end{array}
                  \right)\left(
                    \begin{array}{c}
                      N \\
                      n \\
                    \end{array}
                  \right)
e^{2i(n-m)\phi}(\tan\theta)^{2N-m-n}\nonumber\\
&&e^{-2\kappa
tl^2(n-m)^2}|\alpha-(N-2m)l\rangle\langle\alpha-(N-2n)l|.
\end{eqnarray}
Thus the coherence of the state decays on the time scales
$1/2N^2l^2$. In Fig.\ref{fig4} we show the decoherence effects due
to the cavity damping in the state of the quantum random walker in
terms of Wigner function. As the time progresses from (a) to (d) the
decoherence reduces the quantum interference effects and the state
of the random walker decays to its initial state. In
Fig.\ref{fig4}(a) the Wigner function for the state of the random
walker after $10$ steps using the parameters of Fig.\ref{fig2}(b) is
plotted which is squeezed in $x$ quadrature and centered at
$x\approx-2$. As a result of decoherence due to cavity damping the
quantum interferences start decaying and the Wigner function changes
to the perfect Gaussian shape, Fig.\ref{fig4}(c) centered at $x=Nl$.
Now the field inside the cavity is almost in coherent state and
decays with the cavity damping rate. Further the life time for the
state of the quantum random walker is given by $T_N=T_c/2N^2l^2$
where $T_c=1/\kappa$ is life time for field in the cavity.

\section{conclusions}
In conclusion we have shown a simple possible realization of quantum
random walks using cavity QED. We have proposed homodyne detection
for monitoring the position of the random walker. We have also
discussed the decoherence effects and the time scales at which
quantum nature of random walks survives. As a result of new emerging
technologies various improved cavities are feasible these days
\cite{cavity}, which makes our proposal much interesting and
realistic. Such realization of quantum random walks may be useful
for implementing various algorithms \cite{algorithms} based on
quantum random walks. Finally it should be noted that the
generalizations of the present work to more than one dimensions are
possible.
\appendix*
\section{State of the walker for no measurement on the atomic state}
In this appendix we would like to connect the result (\ref{app})
explicitly to the case of classical random walk. For this purpose we
find the reduced state of the field from (\ref{app}). We also set
$c_1=1,~c_2=0$, then the reduced state of the field $\rho_f$ is
\begin{equation}
\rho_f=\frac{1}{2}\left(|\alpha+\frac{gt}{2}\rangle\langle
\alpha+\frac{gt}{2}|+|\alpha-\frac{gt}{2}\rangle\langle
\alpha-\frac{gt}{2}|\right)
\end{equation}
Clearly the state of the field after the passage of $N$ atoms would
be
\begin{eqnarray}
\rho_f=\left(\frac{1}{2}\right)^N \sum_{m=0}^{N}\left(
                    \begin{array}{c}
                      N \\
                      m \\
                    \end{array}
                  \right)\times~~~~~~~~~~~~~~~~~~~\nonumber\\
|\alpha+\frac{gt}{2}(N-2m)\rangle\langle
\alpha+\frac{gt}{2}(N-2m)|\\
=\left(\frac{1}{2}\right)^N
\sum_{p=-N}^{+N}\frac{N!}{\frac{N-p}{2}!\frac{N+p}{2}!}|\alpha+\frac{gt}{2}p\rangle\langle
\alpha+\frac{gt}{2}p|
\end{eqnarray}
which is reminiscent of the result for classical random walk in the
sense that the weight factor of the state
$|\alpha+\frac{gt}{2}p\rangle\langle \alpha+\frac{gt}{2}p|$ is same
as the probability of finding the walker at the site $p$
\cite{chandra}. It should however be borne in mind that the coherent
states $|\alpha+\frac{gt}{2}p\rangle$ and
$|\alpha+\frac{gt}{2}p'\rangle$ are not orthogonal for $p\neq p'$.

\end{document}